\documentstyle[12pt]{article}
\renewcommand{\theequation}{\arabic{section}.\arabic{equation}}

\newcommand{\be}{\begin{equation}}
\newcommand{\ee}{\end{equation}}
\newcommand{\ba}{\begin{array}}
\newcommand{\ea}{\end{array}}
\newcommand{\bc}{\begin{center}}
\newcommand{\ec}{\end{center}}
\newcommand{\disregard}[1]{{}}
\newcommand{\ti}{\tilde}
\newcommand{\la}{\lambda}
\newcommand{\al}{\alpha}
\newcommand{\am}{\langle\alpha\rangle}
\newcommand{\gam}{\gamma}
\newcommand{\r}{{\cal R}}
\newcommand{\s}{S}
\newcommand{\g}{G}
\newcommand{\ff}{F}
\newcommand{\fipi}{C}
\newcommand{\vf}{{\cal V}}
\newcommand{\hf}{{\cal H}}
\newcommand{\x}{{\bf x}}
\newcommand{\y}{{\bf y}}
\newcommand{\z}{{\bf z}}
\newcommand{\qq}{{\bf r}}
\newcommand{\p}{{\bf p}}
\newcommand{\k}{{\bf k}}

\newcommand{\dt}{\hbox{d}t}
\newcommand{\ct}{\int_{t_0}^{\infty}}
\newcommand{\cd}{{\cal D}}
\newcommand{\ca}{{\cal A}}

\newcommand{\q}{{\bf Q}}
\newcommand{\qm}{\langle Q\rangle}
\newcommand{\m}{{\mu}}
\newcommand{\n}{{\nu}}
\newcommand{\bphi}{{\bf\Phi}}
\newcommand{\bpi}{{\bf\Pi}}
\newcommand{\ds}{\displaystyle}
\newcommand{\demi}{{\ds 1\over\ds 2}}
\newcommand{\quart}{{\ds 1\over\ds 4}}

\newcommand{\prl}[1]{ Phys. Rev. Lett. {\bf #1}}
\newcommand{\prb}[1]{Phys. Rev. {\bf #1}}
\newcommand{\apb}[1]{Ann. of Phys. (N.Y.) {\bf #1} }
\newcommand{\npb}[1]{Nucl. Phys. {\bf #1}}
\title{Variational Multi-Time Green's Functions for Nonequilibrium Quantum
Fields}

\author{
Mohamed Benarous
\thanks{
Perm. address: {University of Chlef, dept. of Physics, B.P. 151, 02000 Chlef, Algeria.}
}
\\
{\it 
Division de Physique Th\'eorique \thanks{Unit\'e 
de Recherche des Universit\'es 
Paris XI et Paris VI associ\'e au C.N.R.S.},
}
\\
{\it Institut de Physique Nucl\'eaire,}
\\
{\it F-91406, Orsay Cedex, France}
}
\date{\today}
\begin{document}
\maketitle
\begin{abstract}

The time-dependent variational principle proposed by Balian and 
V\'en\'eroni is used to provide the best approximation to the
generating functional for multi-time Green's functions of a set
of (bosonic) observables $\q_{\m}$. By suitably restricting the trial spaces, 
the computation of the two-time Green's function, obtained by a
second order expansion in the sources, is considerably simplified.  
This leads to a tractable formalism suited to
quantum fields out of equilibrium. We propose an illustration
on the finite temperature $\bphi^4$-theory in curved space and 
coupled to gravity.
\end{abstract}

Preprint IPNO/TH 97-28     PACS: 05.30.D, 11.15.Tk, 03.65.W

\newpage

\setcounter{section}{0}

\setcounter{equation}{0}
\bc{\section{Introduction}} \ec

In the past few years, a time-dependent variational principle was
proposed \cite{BV85} in order to deal with the question of how
to predict correctly, within an approximation scheme, the expectation
value of a measured observable, when the system is in a mixing of states
and is therefore described by a density operator. The variational principle
has been applied to a variety of quantum processes including heavy ion
reactions\cite{BBFV}, quantum fields out of equilibrium\cite{EJS88}, attempts
to go beyond the gaussian approximation for fermion\cite{Fl89} and
boson systems as well as studies of tunneling through a gaussian
barrier\cite{BM91}.

More recently, a more elaborate version of the variational principle was
developed\cite{BV93} in order to evaluate the generating functional for
multi-time correlation functions in equilibrium and non-equilibrium statistical
mechanics.

The purpose of the present paper is to extend the formalism of \cite{BV93} to
quantum field theory and to make it more transparent in this area. Among the net
advantages of the approach proposed here and in \cite{BV93}, is the ability to
handle in a consistent and non-perturbative way out-of-equilibrium phenomena such 
as the quantum non-equilibrium evolution of the inflaton field(s) in the inflationary
scenario\cite{infl} or that of the quark-gluon plasma\cite{CKMP95,BVH95}.

We will concentrate in this paper on boson fields because of their direct
connections with the previously mentioned problems. Extension to fermion
fields is quite trivial.

The paper is organized as follows: in section 2, we recall some results
concerning the exact determination of multi-time correlation functions
of a set of observables. In section 3, we present the variational action-like
which leads to the desired generating functional as its stationary value.
A peculiarity of the variational principle is the appearance of two variational
objects akin to a density operator and to an observable. This is however
analogous to the bra and ket in ordinary quantum mechanics or to the position
and its conjugate momentum in classical dynamics. In section 4, we select for
the variational objects gaussian operators and derive the corresponding
evolution equations. Being coupled, these equations are quite complicated
because of the mixed boundary conditions inherent in the variational principle.
Fortunately, the computation of the one and two-time correlation functions does
not require the solution of the whole set of equations. Indeed, we show in
section 5 that it is sufficient to expand these equations up to second order in
a set of sources, and this transforms the problem into two initial value
problems.

In order to illustrate the formalism, we choose in section 6 to focus on a
$O(1)$ quantum field theory in Robertson-Walker metric. Furthermore, we
restrict the set of relevant observables (that is, those we wish to calculate
the correlation functions,) to the boson field operator $\bphi (\x)$ and the
composite operator $\bphi (\x)\bphi(\y)$. This allows the determination of the
one-, two-, three- and four-point connected Green's functions of the theory.
These are manifestly finite since the underlying dynamical equations are already
renormalized.

Finally, we present some perspectives and discuss possible improvements of the
approximations involved in this work.

\setcounter{equation}{0}
\bc{\section{Summary of Some Exact Results}}\ec

We wish to determine in this section the exact multi-time correlation
functions of a set of operators, which may be composite, denoted in the
Schr\"odinger picture by $\q_{\m}$ ($\m =1,2\ldots$). Following 
ref.\cite{BV93}, we introduce the generating functional
\be\label{eq1}
W\{\xi\} \equiv -i\,\log{\hbox{Tr}\,A(t_0)D(t_0)},
\ee
where $D(t_0)$ is the exact density operator of the system at some initial 
time $t_0$ (possibly equal to $-\infty$) and $A(t_0)$ is given by
\be\label{eq2}
A(t_0)=T\exp{\left\{i\ct\dt^{'}\sum_{\m}\xi_{\m}(t^{'})
\q_{\m}^H (t^{'})\right\}}.
\ee
In expression (\ref{eq2}), $\xi_{\m} (t^{'})$ are the sources and 
$\q_{\m}^H (t^{'})$ are the Heisenberg representations of the
observables $\q_{\m}$
\be\label{eq3}
\q_{\m}^H (t^{'})=U^+ (t^{'},t_0)\q_{\m} U(t^{'},t_0),
\ee
$U$ being the evolution operator. The index $\m$ may contain both discrete 
indices and space variables that the observables $\q$ may possess.
By expanding (\ref{eq2}) in powers of the sources, we obtain the following 
formal development for the generating functional 
\be\label{eq4}
\ba{rl}
iW\{\xi\} & = \log{\hbox{Tr}\,D(t_0)}+i\ct\dt^{'}\sum_{\m}\xi_{\m}(t^{'})
\langle\q_{\m}^H (t^{'})\rangle \\
& +{\ds i^2\over\ds 2}\ct
\dt^{'}\dt^{''}\sum_{\m ,\n}\xi_{\m}(t^{'})\xi_{\n}(t^{''})
\langle T\left(\bar{\q}_{\m}^H (t^{'})\bar{\q}_{\n}^H (t^{''})\right)
\rangle 
+\ldots ,
\ea
\ee
where $\langle O\rangle =\hbox{Tr}\,OD(t_0)/\hbox{Tr}\,D(t_0)$ and 
$\bar{O}=O-\langle O\rangle$ for any operator $O$. It is now clear
that $W\{\xi\}$ generates the connected Green's functions of the
observables $\q_{\m}$. Indeed, by multiple differentiation with 
respect to the sources, one has for instance
\be\label{eq5}
\langle\q_{\m}^H (t^{'})\rangle ={\ds\delta W\{\xi\}\over\ds
\delta\xi_{\m}(t^{'})}{\Big\vert}_{\xi =0},
\ee
\be\label{eq6}
\langle T\left(\bar{\q}_{\m}^H (t^{'})\bar{\q}_{\n}^H (t^{''})
\right)\rangle \equiv G_{\m\n}^{(2)}(t^{'},t^{''}) =-i
{\ds\delta^2 W\{\xi\}\over\ds
\delta\xi_{\m}(t^{'})\delta\xi_{\n}(t^{''})}{\Big\vert}_{\xi =0}.
\ee

The next step is to define the generating functional for the ''$1-$PI''
Green's functions, that is the effective action $S_{eff}$ as a (multiple)
Legendre transform of $W\{\xi\}$. Upon setting 
\be\label{eq8}
\ba{rl}
\langle\q_{\m}(t^{'})\rangle & \equiv {\ds\delta W\{\xi\}\over\ds
\delta\xi_{\m}(t^{'})} = \langle\q_{\m}^H (t^{'})\rangle 
+ \sum_{n=2}^{\infty} i^{n-1}\ct \dt_1\ldots\dt_n \\
& \sum_{\n_1\ldots\n_{n-1}}\xi_{\n_1}(t_1)\ldots\xi_{\n_{n-1}}(t_{n-1})
G_{\m\n_1\ldots\n_{n-1}}^{(n)}(t^{'},t_1,\ldots t_{n-1}),
\ea
\ee
where
\be\label{eq9}
G_{\n_1\ldots\n_n}^{(n)}(t_1,\ldots t_n )
=\langle T\left(\bar{\q}_{\n_1}^H (t_1 )\ldots
\bar{\q}_{\n_n}^H (t_n )\right)\rangle ,
\ee
we can define the effective action as
\be\label{eq10}
S_{eff}\{\qm\} = W\{\xi\}-\ct\dt^{'}
\sum_{\m}\xi_{\m}(t^{'})\langle\q_{\m}(t^{'})\rangle .
\ee
Now since
\be\label{eq11}
{\ds\delta S_{eff}\{\qm\}\over\ds\delta\langle\q_{\m}(t^{'})\rangle}
=-\xi_{\m}(t^{'}) ,
\ee
the physical theory will be defined by the vanishing of the sources or 
equivalently by the condition
\be\label{eq12}
\langle\q_{\m}(t^{'})\rangle\{\xi =0\} =\langle\q_{\m}^H (t^{'})
\rangle .
\ee

The computation of the $1-$PI Green's functions proceeds by multiple
differentiation of $S_{eff}$ with respect to 
$\langle\q_{\m}(t^{'})\rangle$. But this is a rather difficult task in general 
since it requires the inversion of the functional relation (\ref{eq8}).
However, considerable simplifications occur when one intends to calculate
the two-point $1-$PI function $\Gamma^{(2)}$. Indeed, up to first order in
the sources, the expansion (\ref{eq8}) yields
\be\label{eq13}
\xi_{\m}(t)\simeq -i\ct\dt^{'}\sum_{\n}
\left(\left[G^{(2)}\right]^{-1}\right)_{\m\n} (t,t^{'})
\left(
\langle\q_{\n}(t^{'})\rangle -\langle\q_{\n}^H (t^{'})\rangle
\right) ,
\ee
where one notes that the condition (\ref{eq12}) is preserved. 
Omitting trivial constant terms, the expressions for $W\{\xi\}$ and 
$S_{eff}\{\qm\}$ now read
\be\label{eq14}
W\{\xi\}\simeq\ct\dt^{'}\sum_{\m}
\xi_{\m}(t^{'})\langle\q_{\m}^H (t^{'})\rangle
+{\ds i\over\ds 2}\ct\dt^{'}\dt^{''}\sum_{\m\n}
\xi_{\m}(t^{'}) G_{\m\n}^{(2)}(t^{'},t^{''})\xi_{\n}(t^{''}),
\ee
\be\label{eq15}
S_{eff}\{\qm\}\simeq {\ds i\over\ds 2}\ct\dt^{'}\dt^{''}\sum_{\m\n}
(\langle\q_{\m}\rangle -\langle\q_{\m}^H \rangle )(t^{'})
\Gamma_{\m\n}^{(2)}(t^{'},t^{''})
\langle\q_{\n}\rangle -\langle\q_{\n}^H \rangle )(t^{''}),
\ee
where
\be\label{eq16}
\Gamma_{\m\n}^{(2)}(t^{'},t^{''})\equiv -i {\ds\delta^2 S_{eff}\{\qm\}
\over\ds
\delta\langle\q_{\m}(t^{'})\rangle\delta\langle\q_{\n}(t^{''})\rangle}
{\Big\vert}_{\qm =\langle Q^H \rangle} =
\left(\left[G^{(2)}\right]^{-1}\right)_{\m\n} (t^{'},t^{''}) .
\ee
One should note that, although (\ref{eq14}) and (\ref{eq15}) are truncated,
the expression (\ref{eq16}) is exact since $\Gamma^{(2)}$ is defined at the
''stationary point'' where the sources vanish. 

The determination of higher order equal time Green's functions
proceeds simply by enlarging the set of operators $\q_{\m}$. For instance,
one may take $\q_1 =\q$ (where $\q$ is some simple observable) and 
$\q_2$, $\q_3$,$\ldots$ as successive powers of $\q$. We will see an example
in section 6.

The expressions that we have derived for the correlation functions require 
the exact computation of $W\{\xi\}$ and $S_{eff}\{\qm\}$. But this in turn
needs the evaluation of the exact evolution operator which is a quite 
formidable task in general. In the next section, we present a method first
developed in ref.\cite{BV93} and devised to provide the best approximation
to $W\{\xi\}$ within a variational framework.

\setcounter{equation}{0}
\bc{\section{Variational Formulation}}\ec

We recall in this section the major steps in a variational derivation of
the dynamics for the generating functional $W\{\xi\}$. For a more detailed
analysis, we refer the reader to \cite{BV93} sect. 3. 

The authors of \cite{BV93} introduce the functional (with its two variants)
\be\label{eq300}
\ba{rl}
{\cal I}\{\ca (t);&\cd (t)\} = \hbox{Tr}\,\ca (t_0 )\cd(t_0 )+\\
&+\ct \hbox{Tr}\,
\cd (t)\left\{{\ds\hbox{d}\ca (t)\over\ds\hbox{d}t}-i[\ca (t),H]
+i\ca (t) \sum_{\m} \xi_{\m}(t)\q_{\m}\right\} ,
\ea
\ee
\be\label{eq301}
\ba{rl}
{\cal I}\{\ca (t);&\cd (t)\} = \hbox{Tr}\,\ca (\infty )\cd(\infty )-\\
&-\ct \hbox{Tr}\,
\ca (t)\left\{{\ds\hbox{d}\cd (t)\over\ds\hbox{d}t}+i[H,\cd (t)]
-i\sum_{\m} \xi_{\m}(t)\q_{\m}\cd (t)\right\} .
\ea
\ee
In (\ref{eq300}) and (\ref{eq301}), $\cd (t)$ and $\ca (t)$ are two 
time-dependent trial operators, the latter being subject to the constraint
\be\label{eq302}
\ca (\infty ) =A(\infty ) ={\bf 1} ,
\ee
where $A(t)$ is given by (\ref{eq2}) with $t_0$ replaced by $t$. To account
for the dynamics, the functional $\cal I$ should be made stationary with 
respect to variations of $\cd (t)$ and $\ca (t)$. This leads to the following 
equations
\be\label{eq303}
\hbox{Tr}\,
\delta\cd (t)\left [{\ds\hbox{d}\ca (t)\over\ds\hbox{d}t}-i[\ca (t),H]
+i\ca (t) \sum_{\m} \xi_{\m}(t)\q_{\m}\right ]=0 ,
\ee
\be\label{eq304}
\hbox{Tr}\,
\delta\ca (t)\left [{\ds\hbox{d}\cd (t)\over\ds\hbox{d}t}+i[H,\cd (t)]
-i\sum_{\m} \xi_{\m}(t)\q_{\m}\cd (t)\right ]=0 .
\ee
One should in principle also enforce the optimization of the initial state 
by imposing
\be\label{eq305}
\hbox{Tr}\,
\delta\ca (t_0 )\left [\cd (t_0 )-D(t_0 )\right ]=0 .
\ee
However, since we are not directly interested in this paper in the initial 
state problem, we will suppose that $\cd (t)$ and $D(t_0 )$ belong to the 
same class so that one can impose $\cd (t_0 )=D(t_0 )$ as initial condition.
For a careful study of this problem, see \cite{BV93} sect. 6.

When the variations $\delta\ca (t)$ and $\delta\cd (t)$ are unrestricted, 
one obtains from (\ref{eq303}) and (\ref{eq304})
\be\label{eq306}
{\ds\hbox{d}\ca (t)\over\ds\hbox{d}t}-i[\ca (t),H]
+i\ca (t) \sum_{\m} \xi_{\m}(t)\q_{\m}=0 ,
\ee
\be\label{eq307}
{\ds\hbox{d}\cd (t)\over\ds\hbox{d}t}+i[H,\cd (t)]
-i\sum_{\m} \xi_{\m}(t)\q_{\m}\cd (t)=0 .
\ee
One recognizes in (\ref{eq306}) and (\ref{eq307}) the backward Heisenberg 
equation for $\ca (t)$ and the forward Liouville-von Neumann equation for 
$\cd (t)$ with source terms. The stationary value of the action-like 
(\ref{eq300}) reduces in this case to
\be\label{eq308}
{\cal I}_{\hbox{st}}=\hbox{Tr}\,\ca (t_0 )\cd (t_0 )=\hbox{Tr}\,A(t_0 )D(t_0 ).
\ee
Therefore, the generating functional (\ref{eq1}) writes
\be\label{eq309}
W\{\xi\} =-i\log{{\cal I}_{\hbox{st}}} .
\ee
Hence, one has succeeded in constructing a variational principle which has
the desired quantity as its stationary value.

Now, one can build approximation schemes by choosing for $\cd (t)$ and 
$\ca (t)$ some trial spaces. However, the important relation (\ref{eq309})
will remain true only when $\delta\cd\propto\cd$ is an allowed variation
\cite{BV85,BM91} because the integrand in (\ref{eq300}) vanishes at
the stationary point. This excludes for instance normalized operators.
In addition, when $\delta\ca\propto\ca$ is also allowed, the quantity
$\hbox{Tr}\,\ca (t)\cd (t)$ does not depend on the time $t$, as can be 
seen by adding (\ref{eq303}) and (\ref{eq304}). In this case 
($\delta\cd\propto\cd$, $\delta\ca\propto\ca$), the approximate 
stationary value writes
\be\label{eq310}
{\cal I}_{\hbox{st}}=\exp{iW\{\xi\}}=\hbox{Tr}\,\ca (t)\cd (t)
\quad t\ge t_0 ,
\ee
as in the exact situation. Nevertheless, the dynamics will be of course 
different since it is induced by (\ref{eq303},\ref{eq304}) which lead
in general to non-linear, coupled equations of motion for $\ca (t)$ and 
$\cd (t)$ \cite{BM91,BV93}. Moreover, differentiation of (\ref{eq310})
with respect to $\xi$ yields the interesting result
\be\label{eq311}
{\ds\delta W\{\xi\}\over\ds\delta\xi_{\m}(t)}=
{\ds\hbox{Tr}\,\cd (t)\ca (t)\q_{\m}\over\ds\hbox{Tr}\,\cd (t)\ca (t)} .
\ee
In the following, we shall select for both $\cd (t)$ and $\ca (t)$ trial
spaces that meet with the above restrictions: the boson gaussian operators.

\setcounter{equation}{0}
\bc{\section{Approximate Evolution Equations for Gaussian 
State and Observable}}\ec

Before proceeding further, let us introduce our notation.
In what fol\-lows, $\x$ denotes a $D-$dimen\-sional vector and 
$\int_{\x}=$$\int\hbox{d}^D \x$. In addition, $x$ is the
$(D+1)-$vector $(\x ,t)$. Although the formalism presented here ap\-plies
to quantum field theory, we will use a many-body notation, hiding as much
as pos\-sible space inte\-grals and dis\-crete sums.

Let us first introduce the $2N-$component operator $\alpha (\x )$
in the Schr\"o\-dinger picture
\be\label{eq401}
\ba{rl}
\alpha_j (\x ) & ={\ds 1\over\ds\sqrt{2}}
\left(\bphi_j (\x )+i\bpi_j (\x )\right)
\qquad\quad j=1,2\ldots N,\\
& ={\ds 1\over\ds\sqrt{2}}
\left(\bphi_{j-N} (\x )-i\bpi_{j-N} (\x )\right)
\quad j=N+1,N+2\ldots 2N,
\ea
\ee
where $\bphi (\x)$ is the ($N-$component) boson field operator 
and $\bpi (\x)$ its conjugate momentum. The usual boson commutation relations
write in term of the operator $\alpha$ as
\be\label{eq403}
[\alpha_i (\x ),\alpha_j (\y )]=\tau_{ij}(\x ,\y ),
\ee
where the $2N\times 2N$ matrix $\tau$ has the block form
\be\label{eq4033}
\tau (\x ,\y )=\pmatrix{0 & \delta^D (\x -\y ) \cr -\delta^D (\x -\y ) & 0\cr}
.
\ee

We now choose for both $\cd (t)$ and $\ca (t)$ the class of gaussian
operators in $\alpha$ \cite{BB69}:
\be\label{eq404}
\left\{
\ba{rl}
\cd (t) & = {\cal N}_d \exp{(\la_d\tau\alpha )}
\exp{(\demi\alpha\tau\s_d \alpha )}
,\\
\ca (t) & = {\cal N}_a \exp{(\la_a\tau\alpha )}
\exp{(\demi\alpha\tau\s_a \alpha )}.
\ea\right.
\ee
In (\ref{eq404}), ${\cal N}_d$ and ${\cal N}_a$ are c-numbers, 
$\la_d (\x ,t)$ and $\la_a (\x ,t)$ are $2N-$component
vectors and $\s_d (\x ,\y ,t)$ and $\s_a (\x ,\y ,t)$ are $2N\times 2N$ 
matrices which can be chosen so that $\tau\s_d$ and $\tau\s_a$ are symmetric.

The quantities ${\cal N}_q$, $\la_q$ and $\s_q$ (with $q=d$ or $a$) 
uniquely define the operator to which they refer ($\cd$ or $\ca$). 
It is however more convenient to charecterize these operators by 
${\cal Z}_q (t)$, $\am_q (\x ,t)$ and $\r_q (\x ,\y ,t)$ defined as
\be\label{eq406}
\left\{
\ba{rl}
{\cal Z}_q (t) & =\hbox{Tr}\,{\cal T}_q (t),\\
\langle\alpha_i \rangle_q (\x ,t) & ={\ds\hbox{Tr}\,\alpha_i (\x)
{\cal T}_q (t)\over\ds {\cal Z}_q (t)},\\
({\r_q})_{ij}(\x ,\y ,t) & ={\ds\hbox{Tr}\,(\tau\bar{\alpha})_j (\y)
\bar{\alpha}_i (\x){\cal T}_q (t)\over\ds {\cal Z}_q (t)} ,
\ea\right.
\ee
with ${\cal T}_d =\cd$ and ${\cal T}_a =\ca$. We have introduced in 
(\ref{eq406}) the shifted operators $\bar{\alpha}=\alpha-\am$. 
The relations between the two sets (${\cal N}_q$, $\la_q$, $\s_q$) 
and (${\cal Z}_q$, $\am_q$, $\r_q$) are given in the appendix. From 
now on, we consider (${\cal Z}_d$, $\am_d$, $\r_d$) and (${\cal Z}_a$, 
$\am_a$, $\r_a$) as variational parameters. 
To proceed further, we introduce the mixed operators ${\cal T}_{ad} =\ca\cd$ and
${\cal T}_{da} =\cd\ca$ which are also gaussian operators and can therefore be 
characterized by (${\cal Z}_{ad}$, $\am_{ad}$, $\r_{ad}$) and (${\cal
Z}_{da}$, $\am_{da}$, $\r_{da}$) given by (\ref{eq406}) with $q=ad$ or $da$. 
We give in the appendix the relations between (${\cal Z}_{ad,da}$, $\am_{ad,da}$, 
$\r_{ad,da}$) and (${\cal Z}_{a,d}$, $\am_{a,d}$, $\r_{a,d}$). Furthermore, for 
any operator $O$, we will denote by $\langle O\rangle_q$ (with $q=a$, $d$, 
$ad$ or $da$) its expectation value with respect to $\ca$, $\cd$, ${\cal
T}_{ad}$ or 
${\cal T}_{da}$. Our previous notation anticipated this convention.
Because of the cyclic invariance of the trace, we have 
${\cal Z}_{ad} ={\cal Z}_{da}$. Their common value  will be denoted by $\cal Z$. 
According to equation (\ref{eq310}), it is precisely the quantity of interest 
since we have
\be\label{eq407}
{\cal I}_{\hbox{st}}=\exp{iW\{\xi\}}={\cal Z}(t) .
\ee

The reduced functional (\ref{eq300},\ref{eq301}) now takes the form
\be\label{eq408}
{\cal I}={\cal Z}(t_0 )+\ct {\cal Z}(t)\left\{
{\ds\hbox{d}\log{{\cal Z}}\over\ds\dt}{\Big\vert}_{\cd} 
+i({\cal E}_{ad}-{\cal E}_{da} )+i\sum_{\m}\xi_{\m}\langle\q_{\m}\rangle_{da} 
\right\} ,
\ee
or
\be\label{eq409}
{\cal I}={\cal Z}(\infty )-\ct {\cal Z}(t)\left\{
{\ds\hbox{d}\log{{\cal Z}}\over\ds\dt}{\Big\vert}_{\ca} 
-i({\cal E}_{ad}-{\cal E}_{da} )-i\sum_{\m}\xi_{\m}\langle\q_{\m}\rangle_{da} 
\right\} ,
\ee
where ${\cal E}_{ad}$ and ${\cal E}_{da}$ are the expectation values of $H$ with 
respect to ${\cal T}_{ad}$ and ${\cal T}_{da}$
\be\label{eq410}
{\cal E}_q\equiv \langle H\rangle_q={\ds\hbox{Tr}\,H{\cal T}_q
\over\ds\hbox{Tr}\,{\cal T}_q} ,
\ee
with $q=ad$ or $da$ and ${\ds\hbox{d}\log{{\cal Z}}\over\ds\dt}\vert_{\cd}$ (resp. 
${\ds\hbox{d}\log{{\cal Z}}\over\ds\dt}\vert_{\ca}$) denotes a 
time-differentiation in which the parameters of $\cd$ (resp. $\ca$)
are kept fixed.

The equations of motion for $\r_a$ and $\am_a$ are obtained from
the stationarity conditions of (\ref{eq408}) with respect to $\r_d$
and $\am_d$. A straightforward calculation yields the following equations
\be\label{eq411}
\left\{
\ba{rl}
i{\ds\hbox{d}\r_a\over\ds\dt} & = \r_a \hf_{da}^{'} (1+\r_a )-(1+\r_a )
\hf_{ad} \r_a ,\\
i{\ds\hbox{d}\am_a\over\ds\dt} & = -\r_a\vf_{da}^{'}+(1+\r_a )\vf_{ad} +
i{\ds\hbox{d}\r_a\over\ds\dt}(\am_{ad} -\am_{da} ) .
\ea\right.
\ee 
The stationarity conditions of (\ref{eq409}) with respect to $\r_d$
and $\am_d$ yield the equations of motion for $\r_a$ and $\am_a$:
\be\label{eq412}
\left\{
\ba{rl}
i{\ds\hbox{d}\r_d\over\ds\dt} & = \r_d \hf_{ad} (1+\r_d )-(1+\r_d )
\hf_{da}^{'}\r_d ,\\
i{\ds\hbox{d}\am_d\over\ds\dt} & = -\r_d\vf_{ad}+(1+\r_d )\vf_{da}^{'} -
i{\ds\hbox{d}\r_d\over\ds\dt}(\am_{ad} -\am_{da} ) .
\ea\right.
\ee 
In (\ref{eq411},\ref{eq412}), we have introduced the vector $\vf_{ad}$
and the matrix $\hf_{ad}$ given by
\be\label{eq413}
\left\{
\ba{rl}
(\vf_{ad} )_i (\x ,t) & = 
\left(\tau {\ds\partial {\cal E}_{ad}\over\ds\partial\am_{ad}}\right)_i (\x ,t)=
\sum_{j=1}^{2N} \int_{\y}\tau_{ij}(\x ,\y )
{\ds\partial {\cal E}_{ad}\over\ds\partial (\am_{ad} )_j (\y ,t)} ,\\
(\hf_{ad} )_{ij}(\x ,\y ,t) & = 
-2 \left({\ds\partial {\cal E}_{ad}\over\ds\partial\r_{ad}}\right)_{ij}(\x ,\y ,t)=
-2 {\ds\partial {\cal E}_{ad}\over\ds\partial (\r_{ad} )_{ji}(\y ,\x ,t)} .
\ea\right.
\ee
The primed quantities are obtained from (\ref{eq413}) by replacing the 
index $ad$ by $da$ and the ''energy'' ${\cal E}_{da}$ by
\be\label{eq414}
{\cal E}_{da}^{'} \equiv {\cal
E}_{da}-\sum_{\m}\xi_{\m}\langle\q_{\m}\rangle_{da} .
\ee
This amounts to replace the Hamiltonian $H$ by the source-dependent one
$H^{'} \equiv H-\sum_{\m}\xi_{\m}\q_{\m}$ in the $da-$indexed quantities
as can be seen from (\ref{eq408},\ref{eq409}). Furthermore, since 
${\cal Z}_a$ and ${\cal Z}_d$ present no particular interest, 
we have prefered to omit their equations.

The system (\ref{eq411},\ref{eq412}), which to our knoweledge, has never
been derived elsewhere, requires several comments. First of 
all, the coupling between $\cd$ and $\ca$ occurs via the $ad$ and $da-$indexed
variables. This coupling renders the resolution of the problem
quite complicated because of the mixed boundary conditions which are given 
both at the initial time $t_0$ and at the final time $t_1 =\infty$. Supposing
that the initial density operator is a gaussian operator characterized by
$\r_0$ and $\am_0$, we have
\be\label{eq415}
\r_d (\x ,\y ,t_0 )=\r_0 (\x ,\y ) ,\quad
\am_d (\x ,t_0 )=\am_0 (\x ) .
\ee
On the other hand, the condition (\ref{eq302}) implies that 
${\cal N}_a (\infty )=1$, $\la_a (\x ,\infty )=0$ and 
$\s_a (\x ,\y ,\infty )=0$. Since these two last
con\-ditions can\-not be expres\-sed con\-veniently in terms of $\am_a$ 
and $\r_a$ (see \ref{a6}), it is pre\-ferable to consider the equations
of motion of $\la_a$ and $T_a \equiv \exp{(-\s_a )}$ (with
$T_a (\x ,\y ,\infty )=\delta^D (\x -\y )$)
\be\label{eq416}
\left\{
\ba{rl}
i{\ds\hbox{d} T_a\over\ds\dt} & = T_a \hf_{da}^{'}-\hf_{ad} T_a ,\\
i{\ds\hbox{d}\la_a\over\ds\dt} & = \vf_{ad} -T_a \vf_{da}^{'} -
i{\ds\hbox{d} T_a\over\ds\dt}\am_{da} .
\ea\right.
\ee 

An interesting feature of (\ref{eq411},\ref{eq412} or \ref{eq416}) is that 
the ''contractions'' ($\r_{ad}$, $\am_{ad}$) and ($\r_{da}$, $\am_{da}$) associated
with ${\cal T}_{ad}$ and ${\cal T}_{da}$ satisfy the (formally) simple
equations
\be\label{eq417}
\left\{
\ba{rl}
i{\ds\hbox{d}\r_{ad}\over\ds\dt} & = [\r_{ad} ,\hf_{ad} ]\\
i{\ds\hbox{d}\am_{ad}\over\ds\dt} & = \vf_{ad}
\ea\right.
,\quad
\left\{
\ba{rl}
i{\ds\hbox{d}\r_{da}\over\ds\dt} & = [\r_{da} ,\hf_{da}^{'}]\\
i{\ds\hbox{d}\am_{da}\over\ds\dt} & = \vf_{da}^{'}
\ea\right.
.
\ee 
It is important to note here that these equations are not sufficient because 
of the mixed boundary conditions which cannot be implemented
conveniently for $\am_{ad,da}$ and $\r_{ad,da}$. Notice further that only 
$\am_{da}$ and $\r_{da}$ depend directly on the sources. This dissymetry between
${\cal T}_{ad}$ and ${\cal T}_{da}$ is due to the way we have introduced  the source term in 
(\ref{eq300},\ref{eq301}). Finally, upon using (\ref{eq311}), one obtains
\be\label{eq418}
{\ds\delta W\{\xi\}\over\ds\delta\xi_{\m}(t)}=\langle\q_{\m}\rangle_{da} (t) .
\ee
This result shows that the connected Green's functions are embodied in 
$\langle\q_{\m}\rangle_{da}$ and not in $\langle\q_{\m}\rangle_d$ as one 
would have expected from a first view. Following the exact case, one may 
define the effective action as in section 2, with 
$\langle\q_{\m}(t)\rangle$ (given by (\ref{eq8})) replaced by 
$\langle\q_{\m}\rangle_{da} (t)$. The procedure now is to expand the whole 
set of equations in powers of the sources in order to determine approximate
expressions for (\ref{eq5}), (\ref{eq6}) and more generally for the 
successive terms of (\ref{eq4}). This is the purpose of the next section.

\setcounter{equation}{0}
\bc{\section{Expansion in Powers of the Sources}}\ec

When there are no sources, we have at any time $\ca (t) ={\bf 1}$. 
Therefore, $\r_{ad}^{(0)}=\r_{da}^{(0)}=\r_d^{(0)}\equiv\r^{(0)}$ and
$\am_{ad}^{(0)}=\am_{da}^{(0)}=\am_d^{(0)}\equiv\am^{(0)}$. The equations 
(\ref{eq411},\ref{eq412},\ref{eq417}) become identical and lead to
the following equations
\be\label{eq501}
\left\{
\ba{rl}
i{\ds\hbox{d}\r^{(0)}\over\ds\dt} & = [\r^{(0)} ,\hf^{(0)}]=
-2 \left[\r^{(0)} ,{\ds\partial {\cal E}^{(0)}\over\ds\partial\r^{(0)}}
\right] ,\\
i{\ds\hbox{d}\am^{(0)}\over\ds\dt} & = \vf^{(0)} =\tau
{\ds\partial {\cal E}^{(0)}\over\ds\partial\am^{(0)}} ,
\ea\right.
\ee
which are the Time-Dependent Hartree-Fock-Bogoliubov 
(TDHFB) equations\cite{BM91,B97} submitted to the initial conditions
$\r^{(0)} (\x ,\y ,t_0)=\r_0 (\x ,\y)$ and                 
$\am^{(0)}(\x ,\y ,t_0)=\am_0 (\x ,\y)$. 
These equations are the bosonic counterpart of the well-known TDHF equations 
and belong to the class of the time-dependent mean-field approximation. 
An interesting property, which will be used later, is that the ''von-Neumann 
entropy'' $-\hbox{Tr}\,\cd (t)\log{\cd (t)}$ is conserved. 
This leads to the fact that an initial pure state (\ref{a9}) remains pure
during the evolution.

From expressions (\ref{eq5}) and (\ref{eq418}), one sees that the best 
variational approximation to $\langle\q_{\m}^H (t)\rangle$ within our 
trial spaces is given by
\be\label{eq502}
\langle\q_{\m}\rangle_{da}^{(0)} (t)=\langle\q_{\m}\rangle_d^{(0)} (t)
\equiv\langle\q_{\m}\rangle^{(0)} (t) .
\ee
Hence, to evaluate the expectation value of $\q_{\m}$ at a time $t$, 
one runs the TDHFB equations (\ref{eq501}) from $t_0$ to $t$ and then
computes the quantity (\ref{eq502}) by means of Wick's theorem if necessary.

The next order in the sources allows the determination of the two-point 
connected Green's function $G_{\m\n}^{(2)}$. Indeed, upon using (\ref{eq6})
and (\ref{eq418}), one finds
\be\label{eq503}
G_{\m\n}^{(2)}(t^{'},t^{''})=-i {\ds\delta\langle\q_{\m}\rangle_{da} (t^{'})
\over\ds\delta\xi_{\n} (t^{''})}{\Big\vert}_{\xi =0} .
\ee
Since $\langle\q_{\m}\rangle_{da}$ depends solely on $\am_{da}$ and $\r_{da}$, 
we expand these last variables up to first order in the sources
\be\label{eq504}
\am_{da} =\am^{(0)}+\delta\am_{da} ,\quad
\r_{da} =\r^{(0)} +\delta\r_{da} ,
\ee
with
\be\label{eq505}
\ba{rl}
\delta\am_{da} (\x ,t) & = -i\ct\dt^{'}\sum_{\m}\gam_{\m}(\x ,t,t^{'})
\xi_{\m}(t^{'}) ,\\
\delta\r_{da} (\x ,\y ,t) & = -i\ct\dt^{'}\sum_{\m}\r_{\m}(\x ,\y ,t,t^{'})
\xi_{\m}(t^{'}) .
\ea
\ee
The connected Green's function $G_{\m\n}^{(2)}$ now reads
\be\label{eq506}
G_{\m\n}^{(2)} (t^{'},t^{''})=-
{\ds\partial\langle\q_{\m}\rangle^{(0)}(t^{'})\over\ds\partial\am^{(0)}
(t^{'})}\gam_{\n} (t^{'},t^{''})-
{\ds\partial\langle\q_{\m}\rangle^{(0)}(t^{'})\over\ds\partial\r^{(0)}
(t^{'})}\r_{\n} (t^{'},t^{''}) .
\ee
The equations of motion for $\delta\am_{da}$ and $\delta\r_{da}$ are obtained
by developing (\ref{eq417}) around ($\am^{(0)}$,$\r^{(0)}$). 
In terms of the corrections $\gam_{\m}$ and $\r_{\m}$, they take the form
\be\label{eq507}
\left\{
\ba{rl}
i{\ds\hbox{d}\gam_{\m}\over\ds\dt}(t,t^{'}) & =\tau 
{\ds\partial {\cal F}_{\m}(t,t^{'})\over\ds\partial\am^{(0)}(t)} ,\\
i{\ds\hbox{d}\r_{\m}\over\ds\dt} (t,t^{'}) & = 
[\r_{\m}(t,t^{'}) ,\hf^{(0)}(t)]-2 \left[\r^{(0)}(t), 
{\ds\partial {\cal F}_{\m}(t,t^{'})\over\ds\partial\r^{(0)}(t)}\right] ,
\ea\right.
\ee
where
\be\label{eq508}
{\cal F}_{\m}(t,t^{'})=\vf^{(0)}(t)\tau\gam_{\m}(t,t^{'})-
\demi\hbox{tr}\,\hf^{(0)}(t)\r_{\m}(t,t^{'}) -i\langle\q_{\m}
\rangle^{(0)}(t)\delta (t-t^{'}) .
\ee
The symbol $\hbox{tr}$ in (\ref{eq508}), not to be confused with 
$\hbox{Tr}$, denotes a trace on both the discrete indices and the 
space variables. One has for instance
\be\label{eq5088}
\hbox{tr}\,\hf^{(0)}(t)\r_{\m}(t,t^{'})=\sum_{i,j=1}^{2N}
\int_{\x ,\y} \left(\hf^{(0)}\right)_{ij}(\x ,\y ,t)
\left(\r_{\m}\right)_{ji}(\y ,\x ,t,t^{'}) .
\ee
The equations (\ref{eq507}) are not sufficient since $\am_{da}$ and $\r_{da}$ 
(and hence $\gam_{\m}$ and $\r_{\m}$) are neither known at $t_0$ nor at 
$\infty$. Let us instead linearize the equations (\ref{eq416}) by 
setting $T_a^{(1)}=T_a^{(0)}+\delta T_a\equiv {\bf 1}-\s_a^{(1)}$ 
and $\la_a^{(1)}=\la_a^{(0)}+\delta\la_a\equiv\delta\la_a$, where 
$\s_a^{(1)}$ and $\delta\la_a$ are of first order in the sources.
By using a parametrization similar to (\ref{eq505}), namely
\be\label{eq509}
\ba{rl}
L_a^{(1)}(\x ,t) & \equiv \delta\la_a (\x ,t)-
\int_{\y} \s_a^{(1)} (\x ,\y ,t)\am^{(0)}(\y ,t)\\
& = -i\ct\dt^{'}\sum_{\m}\xi_{\m}(t^{'}) L_{\m} (\x ,t^{'},t) ,\\
\s_a^{(1)} (\x ,\y ,t) & = -i\ct\dt^{'}\sum_{\m}\xi_{\m}(t^{'}) 
\s_{\m}(\x ,\y ,t^{'},t) ,
\ea
\ee
we obtain the following equations for $L_{\m}$ and $\s_{\m}$
\be\label{eq510}
\left\{
\ba{rl}
i{\ds\hbox{d} L_{\m}\over\ds\dt}(t^{'},t) & =\tau 
{\ds\partial {\cal K}_{\m}(t^{'},t)\over\ds\partial\am^{(0)}(t)} ,\\
i{\ds\hbox{d}\s_{\m}\over\ds\dt} (t^{'},t) & = 
-2 {\ds\partial {\cal K}_{\m}(t^{'},t)\over\ds\partial\r^{(0)}(t)} ,
\ea\right.
\ee
where
\be\label{eq511}
{\cal K}_{\m}(t^{'},t)=\vf^{(0)}(t)\tau L_{\m}(t^{'},t)+
\demi\hbox{tr}\,[\r^{(0)}(t),\hf^{(0)}(t)]\s_{\m}(t^{'},t) +i\langle\q_{\m}
\rangle^{(0)}(t)\delta (t-t^{'}) .
\ee
According to the general philosophy, these equations are to be solved
backward in time since the final conditions read
\be\label{eq512}
L_{\m}(\x ,t^{'},\infty ) =0 ,\quad
\s_{\m}(\x ,\y ,t^{'},\infty ) =0 .
\ee
Therefore, for $t^{'}<t$, we have $L_{\m}(t^{'},t)=0$ and 
$\s_{\m}(t^{'},t)=0$. For $t^{'}>t$, $L_{\m}(t^{'},t)$ and 
$\s_{\m}(t^{'},t)$ are solutions of (\ref{eq510}) without 
the $\delta (t-t^{'})$ term in ${\cal K}_{\m}$. The jumps
at $t=t^{'}-0$ read
\be\label{eq513}
\left\{
\ba{rl}
L_{\m}(\x ,t^{'},t^{'}-0) & =-\tau 
{\ds\partial\langle\q_{\m}\rangle^{(0)}(t^{'})\over\ds\partial\am^{(0)}
(\x ,t^{'})} ,\\
\s_{\m}(\x ,\y ,t^{'},t^{'}-0) & = 
-2 {\ds\partial\langle\q_{\m}\rangle^{(0)}(t^{'})
\over\ds\partial\r^{(0)}(\y ,\x ,t^{'})} .
\ea\right.
\ee

We now want to derive an alternative and tractable expression for
$G_{\m\n}^{(2)}$ in terms of $L_{\m}$ and $\s_{\m}$. One can show 
first that the quantity
$$
g_{\m\n}\equiv L_{\m}(t^{'},t)\tau\gam_{\n}(t,t^{''})-\demi\hbox{tr}\,
\s_{\m}(t^{'},t)\r_{\n}(t,t^{''})
$$
does not depend on the time $t$ except for jumps at $t=t^{'}$ and
$t=t^{''}$. Indeed, upon using (\ref{eq507},\ref{eq510}), one finds the
expression
$$
\ba{rl}
{\ds\hbox{d} g_{\m\n}\over\ds\dt} & =
\Big\{ 
{\ds\partial\langle\q_{\m}\rangle^{(0)}(t)\over\ds\partial\am^{(0)}
(t)}\gam_{\n} (t,t^{''})+
{\ds\partial\langle\q_{\m}\rangle^{(0)}(t)\over\ds\partial\r^{(0)}
(t)}\r_{\n} (t,t^{''})
\Big\}\delta (t-t^{'}) \\
& +
\Big\{ 
L_{\m} (t^{'},t){\ds\partial\langle\q_{\n}\rangle^{(0)}(t)\over\ds\partial\am^{(0)}
(t)}-\hbox{tr}\,
\s_{\m}(t^{'},t)\left[\r^{(0)}(t),{\ds\partial\langle\q_{\n}\rangle^{(0)}(t)\over\ds\partial\r^{(0)}
(t)}\right]
\Big\}\delta (t-t^{''}),
\ea
$$
where, by choosing $t^{''}>t^{'}$, it is clear that just the first term
survives the integration over $t$. One recognizes in this remaining term the
expression (\ref{eq506}) for $G_{\m\n}^{(2)}$. Hence, $g_{\m\n}$ can be evaluated
at $t=t_0$ which yields 
\be\label{eq514}
\ba{rl}
G_{\m\n}^{(2)}(t^{'},t^{''}) & =
L_{\m}(t^{'},t_0)\tau\gam_{\n}(t_0 ,t^{''})-\demi\hbox{tr}\,
\s_{\m}(t^{'},t_0 )\r_{\n}(t_0 ,t^{''}) \\
& = L_{\m}(t^{'},t_0)\tau\r_0 L_{\n}(t^{''},t_0 )+\demi\hbox{tr}\,
\s_{\m}(t^{'},t_0 )\r_0 \s_{\n}(t^{''},t_0 )(1+\r_0 ) .
\ea
\ee
The second line in expression (\ref{eq514}) follows from the linearization
of the relations between $(\am_{da} ,\r_{da} )$ and $(\am_a, \am_d ,\r_a, \r_d)$
(\ref{a8}). When evaluated at time $t_0$, the result writes
\be\label{eq5141}
\left\{\ba{rl}
\delta\r_{da} (t_0 )  & = -\r_0 \s_a^{(1)} (t_0 ) (1+\r_0 ) ,\\
\delta\am_{da} (t_0 ) & =  \r_0 L_a^{(1)} (t_0 ),
\ea\right.
\ee
or equivalently
\be\label{eq5142}
\left\{\ba{rl}
\r_{\m}(t_0,t^{''})   & = -\r_0 \s_{\m}(t^{''},t_0) (1+\r_0 ) ,\\
\gam_{\m}(t_0,t^{''}) & =  \r_0 L_{\m}(t^{''},t_0).
\ea\right.
\ee
The anti-causal Green's function is obtained from (\ref{eq514}) by
exchanging ($\mu$,$\n$) and ($t^{'}$,$t^{''}$).

The computation of the two-point Green's function now proceeds 
as follows: one first solves the TDHFB equations (\ref{eq501}) from 
$t_0$ to $t$ in order to determine $\am^{(0)}(t)$ and $\r^{(0)}(t)$.
Then, one runs backward (\ref{eq510}) from $t=t^{'}$ down to $t=t_0$ to obtain
$L_{\m} (t^{'},t_0 )$ and $\s_{\m} (t^{'},t_0 )$. We note in particular
that one is spared solving (\ref{eq507}).

As discussed in \cite{BV93} sect. 4, this procedure reveals several 
interesting features that are missing in the naive application of 
Wick's theorem. Moreover, since $\langle\q_{\m}\rangle_{da} (t)$ and
$\langle\q_{\m}\rangle^{(0)}(t)$ as given by (\ref{eq418}) and 
(\ref{eq502}) are the best second order approximations to 
$\langle\q_{\m}\rangle(t)$ and $\langle\q_{\m}^{H}(t)\rangle$ 
(respectively given by (\ref{eq8}) and (\ref{eq12})), we can use 
(\ref{eq14}) and (\ref{eq15}) to derive directly the expressions
\be\label{eq515}
W\{\xi\}\simeq\ct\dt^{'}\sum_{\m}
\xi_{\m}(t^{'})\langle\q_{\m}\rangle^{(0)}(t^{'})
+{\ds i\over\ds 2}\ct\dt^{'}\dt^{''}\sum_{\m\n}
\xi_{\m}(t^{'}) G_{\m\n}^{(2)}(t^{'},t^{''})\xi_{\n}(t^{''}) ,
\ee
\be\label{eq516}
S_{eff}\{\qm_{da}\}\simeq\frac{i}{2}\ct\dt^{'}\dt^{''}\sum_{\m\n}
\Big(
\langle\q_{\m}\rangle_{da} -\langle\q_{\m}\rangle^{(0)}
\Big)(t^{'})
\Gamma_{\m\n}^{(2)}(t^{'},t^{''})
\Big(
\langle\q_{\n}\rangle_{da} -\langle\q_{\n}\rangle^{(0)}
\Big)(t^{''})
,
\ee
with the relation
\be\label{eq517}
\langle\q_{\m}\rangle_{da} (t)= \langle\q_{\m}\rangle^{(0)}(t)+i\ct\dt^{'}
\sum_{\n}\xi_{\n}(t^{'}) G_{\m\n}^{(2)}(t,t^{'}) .
\ee

\setcounter{equation}{0}
\bc{\section{Illustration}}\ec

Since the TDHFB equations (\ref{eq501}) remain the main ingredient
whatever the set $\{\q_{\m}\}$, they deserve a deeper analysis.
Let us focus on a $O(1)$ quantum field theory in $(D+1)-$dimensional
Robertson-Walker spacetime described by the metric
$\hbox{d}s^2 = \hbox{d}t^2 -a^2 (t)\hbox{d}\x^2$. The Hamiltonian density is
taken to be
\be\label{eq607}
H=\demi a^{-D}(t)\bpi^2 (\x )+a^D (t) V(\bphi ),
\ee
where
\be\label{eq616}
V(\bphi )= \demi a^{-2} \big(\nabla\bphi (\x)\big)^2 +
\demi (m_0^2 +g_0 R)\bphi^2 (\x)+ {\ds\la_0 \over\ds 6}\bphi^4 (\x),
\ee
with the scalar curvature given by (\cite{EJS88,bida})
$R=2D \ddot{a}/a+D(D-1)(\dot{a}/a)^2$.

Upon defining the expectation values
\be\label{eq601}
\left\{\ba{rl}
\phi (\x ,t) &=\langle\bphi (\x )\rangle \\
\pi (\x ,t) &=\langle\bpi (\x )\rangle
\ea\right.,\quad
\left\{\ba{rl}
\g (\x ,\y ,t) &=\langle\bar{\bphi}(\x )\bar{\bphi}(\y )\rangle \\
\ff (\x ,\y ,t) &=\langle\bar{\bpi}(\x )\bar{\bpi}(\y )\rangle \\
\fipi (\x ,\y ,t) &=\langle\bar{\bphi}(\x )\bar{\bpi}(\y )+
\bar{\bpi}(\y )\bar{\bphi}(\x )\rangle
\ea\right. ,
\ee
which are related to $\am$ and $\r$ by means of (\ref{a11}$-$\ref{a12}), the
energy density writes
\be\label{eq608}
{\cal E}=\demi a^{-D}(t)\left(\pi^2 (\x ,t)+\ff(\x ,\x ,t)\right)
+a^D (t)\langle V\rangle (\phi ,\g ).
\ee
For a gaussian density operator, $\langle V\rangle (\phi ,\g )$ can be
computed by means of Wick's theorem which yields
\be\label{eq618}
\ba{rl}
\langle V\rangle &=
\demi \Big(-a^{-2}\Delta_{\x}+m_0^2+g_0 R
+2\la_0\g(\x,\x,t)+{\ds\la_0\over\ds 3}\phi^2(\x,t)\Big)\phi^2(\x,t) \\
&+\demi\Big(-a^{-2}\Delta_{\x}+m_0^2+g_0R+\la_0\g(\x,\x,t)\Big)\g(\x,\x,t).
\ea
\ee

In terms of the Fourier transforms of the new variables
\be\label{eq6101}
\left\{\ba{rl}
\hat{\phi} & = a^{D/2} \phi \\
\hat{\pi}  & = a^{-D/2}\pi +\frac{DH}{2} a^{D/2}\phi
\ea\right. ,\quad
\left\{\ba{rl}
\hat{\g} & = a^D \g \\
\hat{\ff}  & = a^{-D}\ff +(\frac{DH}{2})^2 a^D\g +\frac{DH}{4} 
(\fipi +\ti{\fipi}) \\
\hat{\fipi} & = \fipi +DH a^D \g
\ea\right. ,
\ee
the TDHFB equations take the form
\be\label{eq609}
\left\{\ba{rl}
\dot{\hat{\phi}} & = \phantom{-}\hat{\pi},\\
\dot{\hat{\pi}} & = -\big(\mu^2 -{\ds 4\la_0 \over\ds 3}a^{-D}{\hat{\phi}}^2
\big)\hat{\phi},
\ea\right.
\ee
\be\label{eq621}
\left\{\ba{rl}
\dot{\hat{\g}} & = \hat{\fipi},\\
\dot{\hat{\ff}} & = -(a^{-2}\p^2+\mu^2)\hat{\fipi},\\
\dot{\hat{\fipi}} & = 2\hat{\ff}-2(a^{-2}\p^2+\mu^2)\hat{\g},
\ea\right.
\ee
where we have introduced the time-dependent self-consistent mass $\mu (t)$:
\be\label{eq622}
\mu^2 (t) = m_0^2 +\frac{DH^2}{4}+(g_0 -\quart)R+2\la_0 a^{-D}
({\hat{\phi}}^2+\int_{\k}\hat{\g}(\k,t)),
\ee
$H=\dot{a}/a$ being the Hubble parameter and $\int_{\k}=\int d^D \k /(2\pi)^D$.
In deriving (\ref{eq609}-\ref{eq621}), we have assumed that the field
$\hat{\phi}$ is homogeneous, which is compatible with translation invariance.

The divergences in the $\k-$integral of (\ref{eq622}) are easily removed by
using the adiabatic expansion method\cite{bida,ESS89}. In the limit $D=3$,
this leads to the following prescription
\be\label{eq626}
\left\{\ba{rl}
& {\ds m_R^2\over\ds\la_R} = {\ds m_0^2\over\ds\la_0} ,\\
& {\ds 1\over\ds\la_R} = {\ds 1\over\ds\la_0}+
{\ds 4\over\ds (4\pi)^{\frac{D+1}{2}}}{\ds 1\over\ds 3-D} ,\\
& {\ds g_R -\frac{D-1}{4D}\over\ds\la_R} =
{\ds g_0 -\frac{D-1}{4D}\over\ds\la_0}.
\ea\right.
\ee

Having now established the renormalizability of the TDHFB equations,
we will concentrate on the case where $\q_{\m =1}$ is the $O(1)$ boson
field ope\-rator $\bphi (\x )$ and $\q_{\m =2}$ is the composite operator
$\bphi(\x)\bphi(\y)$. We have therefore to consider for the set $\{\xi_{\m}\}$
a local source $J(\x,t)$ as well as a bilocal one $K(\x,\y,t)$.

According to (\ref{eq509}), $L_{\m}$ are $2-$dimensional vectors and $S_{\m}$
are $2\times 2$ matrices, which, for convenience, we will decompose in the form
\be\label{eq611}
L_{\m}={\ds 1\over\ds\sqrt{2}}
\left(\ba{c}
l_{\m}+ie_{\m} \\
l_{\m}-ie_{\m} \ea\right),
\quad
S_{\m}=\quart\pmatrix{(u_{\m}+v_{\m}) & -(u_{\m}-v_{\m})+2w_{\m}\cr
(u_{\m}-v_{\m})+2w_{\m} & -(u_{\m}+v_{\m})\cr}.
\ee
The evolution equations for these five parameters are obtained from 
(\ref{eq510}-\ref{eq511}). They read
\be\label{eq613}
\left\{\ba{rl}
{\ds\hbox{d}l_{\m}\over\ds\hbox{d}t} & = a^{-D} e_{\m} \\
{\ds\hbox{d}e_{\m}\over\ds\hbox{d}t} & = a^D\Big(
-(-a^{-2}\Delta_{\x}+M^2)l_{\m}
+2\la_R \hat{\phi}(\frac{i}{2}\hat{\fipi}u_{\m}+g_f w_{\m})
\Big)
\ea\right. ,
\ee
\be\label{eq6133}
\left\{\ba{rl}
{\ds\hbox{d}u_{\m}\over\ds\hbox{d}t} & = 2ia^{-D} w_{\m} \\
{\ds\hbox{d}v_{\m}\over\ds\hbox{d}t} & = 2ia^D\Big(
4\la_R \hat{\phi}l_{\m}-(-a^{-2}\Delta_{\x}+M^2+2\la_R g_f)w_{\m}
-i\la_R\hat{\fipi}u_{\m}
\Big) \\
{\ds\hbox{d}w_{\m}\over\ds\hbox{d}t} & = 2ia^D\big(
-a^{-2}\Delta_{\x}+M^2\big)u_{\m}-2ia^{-D}v_{\m}
\ea\right. ,
\ee
with $M^2=m_R^2+g_R R+2\la_R (\hat{\phi}^2+g_f)$ and $g_f$ being the
finite part of $\int_{\k}\hat{\g}(\k,t)$. Their boundary conditions 
(\ref{eq513}) write for $\m =1$
\be\label{eq612}
\left\{\ba{rl}
l_1 (\x,\y,t^{'},t^{'}) & = 0 \\
e_1 (\x,\y,t^{'},t^{'}) & = i\delta^D (\x-\y) \\
u_1 (\x,\y,\z,t^{'},t^{'}) & = 0 \\
v_1 (\x,\y,\z,t^{'},t^{'}) & = 0 \\
w_1 (\x,\y,\z,t^{'},t^{'}) & = 0
\ea\right. ,
\ee
and for $\m =2$
\be\label{eq6122}
\left\{\ba{rl}
l_2 (\x,\y,\z,t^{'},t^{'}) & = 0 \\
e_2 (\x,\y,\z,t^{'},t^{'}) & = i\delta^D (\x-\y)\hat{\phi}(\z,t^{'})+
i\delta^D (\x-\z)\hat{\phi}(\y,t^{'}) \\
u_2 (\x,\y,\z,\qq,t^{'},t^{'}) & = 0 \\
v_2 (\x,\y,\z,\qq,t^{'},t^{'}) & = -2(\delta^D (\x-\qq)\delta^D (\y-\z)+
\delta^D (\x-\z)\delta^D (\y-\qq)) \\
w_2 (\x,\y,\z,\qq,t^{'},t^{'}) & = 0
\ea\right. .
\ee

Hence, as mentioned earlier, the optimization of the $n-$point connected
Green's functions (here $n=2$, $3$ or $4$) proceeds by solving first the
TDHFB equations (\ref{eq609}-\ref{eq621}) forward and then, 
(\ref{eq613}-\ref{eq6133}) backward in time. For instance, one may solve
(\ref{eq613}-\ref{eq6133}) with $\mu =1$ and $\mu =2$. Then, one computes
$G_{11}^{(2)}(t^{'},t^{''})\equiv G_{\Phi\Phi}(\x,\y,t^{'},t^{''})$,
$G_{12}^{(2)}(t^{'},t^{''})\equiv G_{\Phi\Phi\Phi}(\x,\y,\z,t^{'},t^{''})$
and $G_{22}^{(2)}(t^{'},t^{''})\equiv G_{\Phi\Phi\Phi\Phi}(\x,\y,\z,\qq,t^{'},t^{''})$ 
by means of (\ref{eq514}).

At equilibrium, that is for a static solution of the TDHFB equations, one can 
notice that the equations (\ref{eq613}-\ref{eq6133}) can be solved exactly 
for a wide class of metrics. One can therefore check in particular that the 
Green's functions $G_{\m\n}^{(2)}$ do only depend on the time-difference 
$t^{'}-t^{''}$ as expected.  

We can also notice that the two sets (\ref{eq613}) and (\ref{eq6133}) completely 
decouple in the symmetric case $\hat{\phi}=0$. Moreover, the boundary conditions 
(\ref{eq612}) imply that $u_1$, $v_1$ and $w_1$ vanish for all times $t$. 
Therefore, the two-point causal function, given by (see eq.\ref{eq514})
\be\label{eq6141}
\ba{rl}
G_{\Phi\Phi}(\x,\y,t^{'},t^{''})=
&-\int_{\z,\qq}l_1(\x,\z,t^{'},t_0)\ff_0(\z,\qq)l_1(\qq,\y,t^{''},t_0)\\
&-\int_{\z,\qq}e_1(\x,\z,t^{'},t_0)\g_0(\z,\qq)e_1(\qq,\y,t^{''},t_0)\\
&+\demi\int_{\z,\qq}l_1(\x,\z,t^{'},t_0)\ti{\fipi_0} (\z,\qq)e_1(\qq,\y,t^{''},t_0)\\
&+\demi\int_{\z,\qq}e_1(\x,\z,t^{'},t_0)\fipi_0 (\z,\qq)l_1(\qq,\y,t^{''},t_0),
\ea
\ee
where $\ff_0$, $\g_0$ and $\fipi_0$ refer to the initial density matrix $\r_0$,
is obtained by solving just one equation, namely
\be\label{eq614}
\big(
{\ds\hbox{d}^2\over\ds\hbox{d}t^2}+DH{\ds\hbox{d}\over\ds\hbox{d}t}
+(-a^{-2}\Delta_{\x}+M^2)
\big) l_1(t^{'},t)=0.
\ee
Moreover, for $\m =2$, one has only to solve (\ref{eq6133}) since $l_2$ and 
$e_2$ identically vanish.

\setcounter{equation}{0}
\bc{\section{Conclusions and Perspectives}}\ec

We have presented in this work a variational approach suited to the optimization
of multi-time Green's functions of a set of observables. By choosing gaussian
boson operators for the variational objects, we obtained two sets of coupled
non-linear evolution equations with two-time boundary conditions. The resolution
of this type of equations is quite complicated. Fortunately, in order to compute
the one and two-time Green's functions, one must expand these equations up to
second order in a set of sources, and this in turn simplifies drastically the
problem, since it reduces to two initial value problems. The first one, forward
in time, turns out to be the time-dependent mean-field equations and the second,
backward in time, is linear in the variational parameters which precisely
determine the desired Green's functions.

We have illustrated this formalism on a $O(1)$ quantum field theory in a RW
metric. For the set of observables, we have chosen the boson field operator
$\bphi (\x)$ and the composite operator $\bphi (\x)\bphi (\y)$. This allowed 
us to optimize the one-, two-, three- and four-point connected Green's functions 
of the observable $\bphi$.

The extension of the present formalism to fermion fields is quite simple since
it only depends on the way we parametrize the contraction matrix $\r$ (remember
that in this case $\am =0$.) However, the generalization to gauge field theories
still remains an open question. In particular, it is not clear whether one
should introduce the gauge conditions as constraints in the variational
principle (which seems very natural) or one should better enforce these
conditions in the choice of the trial spaces. The latter procedure has in fact
been attempted\cite{KV89} but has several drawbacks.

So far we have only discussed gaussian trial operators. The question that
naturally arises is how one can go beyond the gaussian approximation while
keeping the tractability of the formalism and the consistency of the
approximations. Some promising methods do indeed exist in the literature.
Among them, we can quote in particular the ''post-gaussian''
approximation\cite{Fl89,BM91} which consists of an expansion in cumulants 
around a gaussian density operator, or the so-called non-gaussian
calculation\cite{PRI89}. More recently, a new method, based on the
background field method, has been developed\cite{Y97}.

We are grateful to M. V\'en\'eroni, H. Flocard and D. Vautherin for 
fruitful discussions. 
We are particularly indebted to C. Martin for her valuable comments and a
careful reading of the manuscript. We would like to thank the members of the 
Division de Physique Th\'eorique, Institut de Physique Nucl\'eaire (IPN), 
Orsay-France, where part of this work has been done.

In the final elaboration of this work, we received a paper\cite{M95}
dealing with the same approach. We thank C. Martin for sending us the
manuscript.

\newpage

\setcounter{equation}{0}
\renewcommand{\theequation}{A.\arabic{equation}}
{\Large{\bf Appendix}}

For the sake of completeness, we recall in this appendix some useful
properties of bosonic gaussian operators of the form:
\be\label{a1}
{\cal T}={\cal N}\exp{(\la\tau\al)}\exp{(\demi\al\tau S\al)}.
\ee
In (\ref{a1}), $\cal N$ is a c-number, $\la (\x,t)$ a $2N-$component vector
and $S(\x,\y,t)$ is a $2N\times 2N$ symplectic matrix. The ($2N\times 2N$)
matrix $\tau$ is defined as
\be\label{a2}
\tau (\x,\y)=\pmatrix{0&\delta^D (\x-\y)\cr -\delta^D (\x-\y)&0\cr},
\ee
and $\al (\x)$ is the $2N-$component boson operator in the Schr\"odinger
picture
\be\label{a3}
\al_j (\x)=
{\ds 1\over\ds\sqrt{2}}
\left\{\ba{rl}
&\bphi_j (\x)+i\bpi_j (\x)\quad\quad\quad j=1,2\ldots N\\
&\bphi_{j-N}(\x)-i\bpi_{j-N}(\x)\quad j=N+1,\ldots 2N
\ea\right.
\ee
obeying the usual commutation relations
\be\label{a4}
[\al_i(\x) ,\, \al_j(\y)]=\tau_{ij}(\x,\y).
\ee
We are adopting in (\ref{a1}) compact notations for discrete sums and space
integrals. For instance
\be\label{a5}
\ba{rl}
&\la\tau\al =\sum_{i,j=1}^{2N}\int_{\x,\y}\la_i(\x,t)\tau_{ij}(\x,\y)\al_j (\y),
\\
&\al\tau S\al =\sum_{i,j,k=1}^{2N}\int_{\x,\y,\z}\al_i(\x)\tau_{ij}(\x,\y )
S_{jk}(\y,\z,t)\al_k(\z).
\ea
\ee

As mentioned in the text, it is more convenient to work with the set of three
parameters consisting of the ''partition function'' $\cal Z$, the vector $\am$
and the contraction matrix $\r$. These are given by
\be\label{a6}
\left\{\ba{rl}
{\cal Z}\equiv & \hbox{Tr}\,{\cal T}={\cal N}\exp{(\demi\la\tau\r\la)}
\sqrt{\det{\sigma\r}} \\
\am\equiv & \hbox{Tr}\,{\cal T}\al/{\cal Z}={\ds 1\over\ds {1-\exp{(-S)}}}
\la ={\ds 1\over\ds 1-T}\la \\
\r \equiv & \hbox{Tr}\,\big({\cal T}\tau\bar{\al}\ti{\bar{\al}}\big)/{\cal Z}
={\ds 1\over\ds{\exp{(S)}-1}}={\ds T\over\ds 1-T}
\ea\right. ,
\ee
where $T=e^{-S}$, $\bar{\al}=\al -\am$ and
\be\label{a7}
\sigma (\x,\y)=\pmatrix{0&\delta^D (\x-\y)\cr \delta^D (\x-\y)&0\cr}.
\ee

The product of two gaussian operators ${\cal T}={\cal T}_{1}{\cal T}_{2}$
of the form (\ref{a1}) with parameters ${\cal Z}_1$, $\am_1$, $\r_1$ and
${\cal Z}_2$, $\am_2$, $\r_2$, is also a gaussian operator of the form
(\ref{a1}) with parameters ${\cal Z}$, $\am$ and $\r$. These can be written
\be\label{a8}
\left\{\ba{rl}
{\cal Z} = & {\cal Z}_1 {\cal Z}_2 
\exp{\big\{\demi (\am_1-\am_2)\tau{\cal P}(\am_1-\am_2)\big\}}
\sqrt{\det{\sigma{\cal P}}} \\
\am = & \r_1{\cal P}\am_2+(\r_2+1){\cal P}\am_1 \\
\r = & \r_1 {\cal P}\r_2
\ea\right. ,
\ee
with ${\cal P}=(1+\r_1+\r_2)^{-1}$. As an immediate consequence, a projector
on coherent states (that is a pure state density matrix):
${\cal T}^2={\cal Z}{\cal T}$ satisfies the property
\be\label{a9}
\r^2 +\r =0 .
\ee

The expressions of $\am$ and $\r$ become more intuitive in terms of the
usual averages defined as ($i,j=1,\ldots N$)
\be\label{a10}
\left\{\ba{rl}
\phi_i (\x,t) &=\langle\bphi_i(\x)\rangle \\
\pi_i  (\x,t) &=\langle\bpi_i (\x)\rangle
\ea\right.,\quad
\left\{\ba{rl}
\g_{ij}(\x,\y,t) &=\langle\bar{\bphi}_i(\x)\bar{\bphi}_j(\y)\rangle \\
\ff_{ij}(\x,\y,t) &=\langle\bar{\bpi}_i(\x)\bar{\bpi}_j(\y)\rangle \\
\fipi_{ij}(\x,\y,t) &=\langle\bar{\bphi}_i(\x)\bar{\bpi}_j(\y)+
\bar{\bpi}_j(\y)\bar{\bphi}_i(\x)\rangle
\ea\right. ,
\ee
with $\bar{Q}=Q-\langle Q\rangle$. Indeed, with the help of the canonical
transformations (\ref{a3}), one gets easily
\be\label{a11}
\am ={\ds 1\over\ds\sqrt{2}}\left(\ba{c}
\phi +i\pi \\
\phi -i\pi\ea\right),
\ee
and
\be\label{a12}
\r =\demi\pmatrix{
\g+\ff-\frac{i}{2}(\fipi -\ti{\fipi})&
-(\g-\ff)-\frac{i}{2}(\fipi +\ti{\fipi})\cr
(\g-\ff)-\frac{i}{2}(\fipi +\ti{\fipi})&
-(\g+\ff)-\frac{i}{2}(\fipi -\ti{\fipi})\cr}.
\ee

\end{document}